\documentclass[aps,twocolumn,epsfig,eqsecnum]{revtex4}
\newcommand\be{\begin{eqnarray}}
\newcommand\ee{\end{eqnarray}}
\newcommand\ba{\begin{array}}
\newcommand\ea{\end{array}}

\def\r{\rangle}
\def\l{\langle}
\def\T{{\rm Tr}}

\def\cH{{\cal H}}
\def\cS{{\cal S}}
\def\cP{{\cal P}}

\def\cI{{\cal I}}

\def\cT{{\cal T}}

\def\cF{{\cal F}}
\def\cU{{\cal U}}
\def\cE{{\cal E}}

\def\cD{{\cal D}}

\begin{document}

\title{Quantum process tomography: the role of initial correlations}
\author{M\'ario Ziman$^{1,2,3}$}
\address{
$^{1}$Research Center for Quantum Information, Slovak Academy of Sciences,
D\'ubravsk\'a cesta 9, 845 11 Bratislava, Slovakia \\
$^{2}$ Faculty of Informatics, Masaryk University, Botanick\'a 68a,
602 00 Brno, Czech Republic\\
$^{3}$ {\em Quniverse}, L{\'\i}\v{s}\v{c}ie \'{u}dolie 116, 841 04
Bratislava, Slovakia
}
\begin{abstract}
We address the problem of quantum process tomography 
with the preparators producing states correlated with 
the environmental degrees of freedom that play role in the system-environment
interactions. We discuss the physical situations, in which 
the dynamics is described by nonlinear, or noncompletely 
positive transformations. 
In particular, we show that arbitrary mapping 
$\varrho_{\rm in}\to\varrho_{\rm out}$ can be realized
by using appropriate set of preparators and 
applying the unitary operation SWAP. The experimental
``realization'' of perfect NOT operation is presented.
We address the problem of the verification of the compatibility of the
preparator devices with the estimating process.
The evolution map describing the dynamics in arbitrary time interval is known
not to be completely positive, but still linear. The tomography and general
properties of these maps are discussed.

\end{abstract}
\maketitle

\section{Motivation}
The postulates of quantum theory require that the 
dynamics of isolated quantum systems is driven by 
Schr\"odinger equation \cite{schrodinger,perez}, 
i.e. for each time interval the evolution
is described by a unitary transformation. However, 
for open quantum systems the situation is different 
\cite{davies,nielsen} and under certain assumptions 
the evolution is described as an one-parametric sequence of
completely positive tracepreserving linear maps (quantum channels) 
$\cE_t$. These mappings describe the state dynamics
for arbitrary time interval $(0,t)$, however for general 
time intervals $(t_1,t_2)$ the state transformations
$\cE_{t_1,t_2}:\varrho_{t_1}\to\varrho_{t_2}$ do not
necessarily possess the above property of complete positivity. 
The aim of this paper is to analyze the cases, in which the 
description of quantum dynamics is not completely positive, or 
even not linear. One of the discussed problems will be the question
of properties of the evolution map $\cE_{t_1,t_2}$ for intermediate time 
intervals for general dynamics of open system governed by sequence 
$\cE_t$. An important exception is if the sequence 
$\cE_t$ fulfills the semigroup property, 
i.e. $\cE_{t+s}=\cE_t\cE_s$ for all $t,s\ge 0$. In this
case for each intermediate time interval the dynamics is linear
and completely positive. 

The lack of complete positivity and linearity for state
transformations is usually interpreted as an 
unphysical property, i.e. these operations cannot 
be physically realized. But still an optimal physical approximation 
of several physically impossible processes is of great 
importance. Typical examples are quantum NOT operation \cite{buzek_werner}, 
quantum copy operation \cite{zurek,gisin}, etc. These processes
violate the rules of quantum dynamics. However, we will see under 
which circumstances and in which sense even such 
unphysical transformations can be observed in our labs. 

Quantum process tomography is a particular goal
of quantum experiments. Several strategies how to gather valuable
experimental data for this task and methods how to correctly 
proceed such data are known 
\cite{chuang,paris,fiurasek,sacchi,buzek_drobny,ziman,lidar}.
The failure of the direct (inverse) estimation methods that could result
in an unphysical map \cite{paris,ziman}, is usually corrected by usage of 
more sophisticated statistical tools such as
maximum likelihood \cite{fiurasek,sacchi}, 
or Bayesian statistics \cite{buzek_drobny}. These methods 
are ``forced'' to lead to a correct quantum channel. We used to say that
the failure of the direct estimation schemes follows from the
finiteness of the measured statistical sample, i.e.
the observed frequencies do not correspond to
theoretically allowed probabilities and consequently, they do
not correspond to some completely positive tracepreserving 
linear map. Here we shall address the question, under which circumstances
such ``unphysicality'' can be due to imperfect 
(potentially correlated) preparators.

\section{Nonideal preparations}

The usual picture of open quantum system dynamics is based on
three assumptions: i) the physical object under
consideration is a part of some larger system that is isolated, i.e.
its evolution is unitary, ii) initial state of the object 
and the environment is factorized, and iii) the state of the environment
is independent of the state of the system. Under such conditions the
resulting dynamics is completely positive and linear.
The question is whether the unphysical maps obtained as a result
of direct estimation can be interpreted in this picture provided
that we relax the last two conditions, i.e. the initial state
is potentially correlated, or the state of the environment depends on
the system state, or both. This question, namely how the initial 
correlations affect the dynamics of open system, 
has been already studied by several
authors \cite{pechukas,stelmo,jordan,shaji,shaji_dis,carteret}. Authors in 
\cite{jordan,shaji_dis,carteret} analyze this problem and 
propose the mathematical tools how to mathematically describe 
such extended evolution maps. 

A general state of bipartite system can be written in the following form
\be
\varrho_{AB}=\varrho_A\otimes\varrho_B
+\sum_{jk}\Gamma_{jk}\Lambda_j^{A}\otimes\Lambda_k^{B}
\ee
where $\Lambda_0^X=\frac{1}{{\rm dim}\cH_X}I$, 
$\T\Lambda_j^X\Lambda_k^X=\delta_{jk}$ with $X=A,B$, 
$j=0,1,\dots,{\rm dim}\cH_A-1$ and $k=0,1,\dots,{\rm dim}\cH_B-1$.
Coefficients $\Gamma_{jk}=\l\Lambda_j^A\otimes\Lambda_k^B\r_{\varrho_{AB}}-
\l\Lambda_j^A\r_{\varrho_A}\l\Lambda_k^B\r_{\varrho_B}$
form the so-called correlation matrix. 
The evolution $\cE$ of the subsystem $A$ is 
a composition of the following three maps: 1) preparation map \cite{note1} 
$\cP:\cS_A\to \cS_{AB}$ ($\cS_X$ stands for the set of quantum states
of the system $X$) satisfying the property $\T_B\cP[\varrho_A]=\varrho_A$,
2) isolated dynamics $\cU:\cS_{AB}\to\cS_{AB}$, i.e. 
$\varrho^\prime_{AB}=\cU[\varrho_{AB}]=U\varrho_{AB}U^\dagger$,
and 3) partial trace $\cT_B:\cS_{AB}\to\cS_{A}$. The last two mappings
are linear and completely positive. Moreover, both of these two features
are preserved under the composition of mappings. That is, the only
source of ``nonphysicality'' is the preparation map $\cP$. 
One can show \cite{pechukas} that the linearity 
of the resulting dynamical map 
$\cE=\cT_B\circ\cU\circ\cP$ requires $\cP$ be of the 
form $\cP[\varrho_A]=\varrho_A\otimes\xi_B$ with $\xi_B$ arbitrary, but 
fixed. In such case the linearity and complete positivity of $\cE$ holds.

In \cite{carteret} authors studied different types of
preparation maps and define the notion of an accessible map.
The transformation is called accessible if it can be 
written as a composition of the preparation,
unitary transformation and partial trace \cite{access_note}. 
If one allows arbitrary 
initial correlations, then the transformation is composed of
two terms \cite{stelmo}
\be
\varrho_A^\prime=
\sum_{\mu\nu} A_{\mu\nu}\varrho A_{\mu\nu}^\dagger +
\sum_{jk}\Gamma_{jk} \sum_\mu \l\mu|\Lambda_j^A\otimes\Lambda_k^B|\mu\r \, .
\ee
where the operators $A_{\mu\nu}=\l\mu| \sqrt{p_\nu}U|\nu\r$ depends
on $\varrho_B$, because $|\mu\r$ are eigenvectors of 
the operator $\varrho_B$. That is, even if we put $\Gamma=0$, the
transformation $\varrho_A\to\varrho_A^\prime$ is still not necessarily
described by some proper quantum channel, 
because the choice of $\varrho_B$ specifying the 
preparation map $\cP$ can depend on $\varrho_A$. 
Consider an arbitrary state transformation $\varrho_{in}\to\varrho_{out}$.
Let us define a preparation in the following way
$\cP[\varrho_{in}]=\varrho_{in}\otimes\varrho_{out}$. Next apply
the SWAP operation (this is a unitary transformation) to obtain
$U_{\rm SWAP}(\varrho_{in}\otimes\varrho_{out})U^\dagger_{\rm SWAP}=
\varrho_{out}\otimes\varrho_{in}$. After performing the partial trace
we obtain the required state transformation $\varrho_{in}\to\varrho_{out}$.
Moreover, we did not use any correlation (quantum, or classical) 
in our preparation at all. Of course, this construction is a 
bit artificial, but nevertheless it shows
that arbitrary state transformation $\varrho_{in}\to\varrho_{out}$ 
is in principle accessible, i.e. can be written as $\cT_B\circ\cU\circ\cP$.
In order to avoid such ``artificial'' realizations of any map we 
need to pose some well-motivated physical conditions. In the 
Ref.\cite{jordan,shaji_dis,carteret} the authors restrict themselves
to linear preparation maps. In what follows we will
analyze two experimental situations in which a preparation map
is naturally defined and the extended dynamics 
can be studied. 

Before we get further let us mention one very important
implication of the fact that arbitrary channel is accessible.
In a sense this statement is very positive, because
whenever we find out in our experiments some strong evidence that the dynamics
is not linear, or not completely positive, we cannot automatically 
conclude that the quantum theory is not correct. The observed ``unphysicality''
can be still interpreted as the problem of devices called preparators 
that produce states correlated to 
degrees of freedom relevant for the subsequent
system evolution.  Without considering the dynamics the potential 
correlations are not interesting and from kinematic point of view they 
are irrelevant. However, these dynamical aspects can be used to  
differentiate between otherwise kinematically equivalent preparators. 
From this point of view the ``nonphysicality'' means that 
the preparators are not independent of the environmental degrees 
of freedom that do take a role in the dynamics. 

As an example consider now the realization of the perfect
NOT operation realized on pure states $|\psi\r\to|\psi_\perp\r$.
Let us assume that the preparation of the spin-$\frac{1}{2}$ 
pure state is performed by a postselection after Stern-Gerlach measurement.
In this way we can prepare any pure quantum state and mixtures can be
obtained by mixing the pure state preparations. Kinematically
this is completely correct preparation procedure of arbitrary qubit 
state. Imagine a situation that the spin is entangled with another 
spin such that together they are described by the singlet,
i.e. $|\Psi\r=\frac{1}{\sqrt{2}}(|\psi\r|\psi_\perp\r-|\psi_\perp\r|\psi\r)$.
Reading the outcome tells us perfectly which state 
$|\psi\r$ we prepared, but the measurement affects 
also the state of the second spin, which is described by the state
$|\psi_\perp\r$. If the unknown device internally just swaps
these two spins, than we find out that the device performs 
the transformation $|\psi\r\to|\psi_\perp\r$, 
i.e. the perfect NOT operation. Experimenter
using such preparations is not aware of the initial correlations and
therefore he would conclude that the unknown device performs 
perfect quantum NOT operation. This conclusion 
is not wrong and indeed experimenter can prove that 
this device performs NOT operation, but only for specific 
preparation procedures. If he would use different state preparators 
(i.e. kinematically equivalent to the previous ones), 
he will very soon find some contradiction.

\section{Process tomography}

The ``nonphysicality'' is not an exception in 
process tomography using the direct estimation schemes \cite{nonphysicality}. 
In such schemes we usually measure a collection of assignments 
$\varrho_j\to\varrho_j^\prime$ for linearly independent states $\{\varrho_j\}$.
Using the linearity of quantum channels these assignments provide
us with sufficient information to complete the reconstruction task.
In other words: each state $\varrho$ can be written as a linear combination
of states $\varrho_j$, i.e. $\varrho=\sum_j a_j\varrho_j$ 
($a_j$ are arbitrary). Therefore the transformation of $\varrho$
is determined by the set of measured assignments.
However, quite often the resulting map
is not completely positive. Even in cases when all $\varrho_j,\varrho_j^\prime$
are proper quantum states. A reason could be really only the usage
of finite data sample, i.e. the statistics is really small 
to conclude something about actual probabilities, mean values, and states.
This line of arguments leads us to the usage of sophisticated
statistical techniques (maximum likelihood, Bayesian approach, etc.).
These statistical methods are modified for quantum tomography purposes 
in a way that they are essentially 
forced to guarantee a physically valid result 
(a completely positive map in quantum process tomography)
even in situations when the direct inverse schemes fail.
However, as we have just explained, this lack of complete positivity can
be due to presence of correlations in our preparators as well.
If this is the case it is really not easy to say which of the
preparators are not perfect, i.e. which of them represent the source of
nonphysicality. It could happen that only one of them is imperfect, 
or all of them are imperfect.

Without any insight into the physics behind the preparation 
process, the measured data do not contain any information about the
origins and form of the preparation map. As we have seen there is always
a trivial example using the SWAP operation that can be used to interpret
arbitrary result. It is important to say here, that even if the
``linearization'' of assignments gives a correct quantum channel, 
it does not mean that our preparators are perfect. 
We should have in mind that the linearity is not
tested, but only used as the theoretical tool to accomplish 
the process estimation. In order to be sure about linearity 
one really needs to test the action
of the channel on preparators of any quantum state. 
In this sense the specification of the channel is always only a hypothesis
and for specific (imperfect) preparators we can find the channel
to be ``unphysical''. Important point is that the usual notion of quantum
channel has a good meaning only for properly prepared input states, i.e.
for compatible preparators. 

We said that using only the measured data we have only very partial
information about the real physics of the process. There are many
possible unitary representations of the observed assignments. One of the 
option is to say that a collection of used preparators was not perfect
and conclude that the process estimation is not possible.
Another approach is to apply some techniques 
of incomplete process tomography \cite{ziman06}
to estimate the unitary map acting on the system  
with Hilbert space $\cH\otimes\cH_{\rm env}$, where $\cH_{\rm env}$
is arbitrary. However this is indeed a difficult 
task, because we need to deal with data that do not contain complete 
information about the inputs as well as about the outputs of the channel, i.e.
the assignments are not completely known. The method that can be used 
in such situations is called principle of maximum entropy \cite{jaynes}. 
However, these issues are beyond the scope of this paper.

\section{Compatibility of preparator and tested process}

To be sure that the channel estimation gives a physical result
one needs to use the collection of ``good'' preparators producing
linearly independent test states. ``Good'' in a sense that whatever 
degrees of freedom enters the preparation
process, these are irrelevant for the channel realization. 
We will say that such preparators are compatible with the channel realization.
Let us note that the condition of producing a factorized state is not 
sufficient, i.e. even pure state preparators are not automatically free
of imperfections. This follows from the example with SWAP operation,
where no correlations in preparator process are used at all, but any 
transformation can be realized. We see that important question is: 
how to test the quality of the preparator, or better to say, how
to test the compatibility of the preparator and the quantum process?

The motivation for the scheme we are going to use 
comes from the preparation process used in real experiments 
\cite{boulant,weinstein,wunderlich,howard}. 
In some cases the preparation of different
states is done by exploiting quantum processing, i.e.
transforming the known state by a known transformation to obtain
a new state. In particular, let us assume we have a preparator
that produces system in a state $\varrho$. Applying unitary rotations
$U_j$ we are able to prepare states $\varrho_j=U_j\varrho U_j^\dagger$
that can be used to test the properties of an unknown 
quantum channel. Our aim is to test the compatibility of the original 
preparator and some unknown device (black box). 
Except the case of $\varrho=\frac{1}{d}I$ the unitary
processing produce sufficiently many linearly independent states
to perform the complete process tomography. This procedure, of course, 
requires perfect realization and control of unitary transformations $U_j$.
Moreover, these unitaries must be already ``compatible'' with the preparator.
In some sense we are cheating a bit here, because we are going to
test the compatibility of the preparator with a given device and
we already assume we have devices (performing unitaries) 
compatible with the preparator. However, as we said this
is quite usual procedure how to prepare different
states in many experiments. Therefore, let us assume that we 
indeed have such compatible devices. Then the described setting 
can be used to test the quality 
of the single preparator with respect to the realization of the 
unknown channel $\cE$. Important point is that these "preparing operations''
are independent of the original preparator, i.e. they do not introduce
the ``unphysicality'' and the only source of ``unphysicality''
is the preparator of $\varrho$. Let us note that there is no
need for preparing operations to be unitary, but in general,
they must be linear and completely positive in order not to introduce
extra sources of ``unphysicality''.

In what follows we will study the properties of the preparation map
given the described model of the experiment consisting of
a preparator, an unknown black box $\cE$ 
and a set of preparing operations $\{\Phi\}$
(not necessarily unitary ones). Consider that the 
original preparator (the one we want to test with 
respect to the channel action) produces states 
$\omega=\varrho\otimes\varrho_B+\Gamma$, where 
$\Gamma=\sum_{jk}\Gamma_{jk}\Lambda_j\otimes\Lambda_k$ stands for the 
correlations. After fixing the set of preparing operations $\{\Phi\}$
the preparation $\cP$ is, in general, defined as follows
\be
\cP:\varrho_\Phi \mapsto \omega_\Phi = \Phi\otimes\cI[\omega]
=\varrho_\Phi\otimes\varrho_B+\Gamma^\prime
\ee
with $\Gamma^\prime=\Phi\otimes\cI[\Gamma]=
\sum_{jk}\Gamma_{jk}^\prime\Lambda_j\otimes\Lambda_k$ 
($\Gamma_{jk}^\prime=\sum_{l} \Phi_{jl}\Gamma_{lk}$, $\Phi_{jl}=
\T(\Lambda_j\Phi[\Lambda_k])$) and $\varrho_\Phi=\Phi[\varrho]$.
The possible preparation mappings $\cP$ are represented by subsets 
of completely positive tracepreserving linear maps $\Phi$ that transform
the given state $\varrho$ into arbitrary state in the domain of $\cP$.
The simple example with the SWAP gate is excluded/trivial in this case, 
because it generates only contractions into the fixed state $\varrho_B$.
Moreover, the correlations induced in the preparation process
are in some sense fixed by the state $\omega$, i.e. 
by the original state preparator that we are testing 
(together with the channel reconstruction).

The unitary evolution $U=\sum_{\mu,\nu}A_{\mu\nu}\otimes|\mu\r\l\nu|$
induces the state transformation
\be
\varrho_\Phi\to\varrho^\prime_\Phi=
\sum_{\mu,\nu} \lambda_\nu A_{\mu\nu}\varrho_\Phi A^\dagger_{\mu\nu} 
+\sum_{j,k,\mu,\nu,\nu^\prime} 
\Gamma_{jk}^\prime [\Lambda_k]_{\nu\nu^\prime} 
A_{\mu\nu}\Lambda_j A_{\mu\nu^\prime}^\dagger\, ,
\ee
where we used that $\varrho_B=\sum_\nu \lambda_\nu |\nu\r\l\nu|$, i.e.
$\{|\nu\r\}$ are eigenvectors of $\varrho_B$. Since $\varrho_B$ is 
fixed for any operation $\Phi$ we see that the first term is 
independent of $\Phi$, i.e. it is independent of the preparation
map $\cP$. Consequently, the unknown 
transformation $\cE$ describing the device can be written as 
the sum of linear completely positive map $\cF$ 
($\cF[\varrho]=\sum_{\mu\nu} F_{\mu\nu}\varrho F_{\mu\nu}^\dagger$
with $F_{\mu\nu}=\sqrt{p_\nu}A_{\mu\nu}$) 
and some traceless operator \cite{carteret} 
$\Xi_\varrho=\sum_{j,k,\mu,\nu,\nu^\prime} 
\Gamma_{jk}^\prime [\Lambda_k]_{\nu\nu^\prime} 
A_{\mu\nu}\Lambda_j A_{\mu\nu^\prime}^\dagger$, i.e.
$\varrho\to\varrho^\prime=\cE[\varrho]=\cF[\varrho]+\Xi_\varrho$. 
First part is irrelevant of the correlations, 
but the second part
can be even nonlinear. The properties of $\Xi_\varrho$ depends completely
on the choice of the set of preparing operations $\Phi$, i.e.
choice of preparation mapping $\cP$. In special cases it can be linear
and not factorized, but then it cannot be defined on the whole 
state space. The linear case was analyzed in Ref.\cite{carteret}), 
in which the sufficient conditions for its existence 
(in terms of properties of $\Xi_\varrho$) were formulated.
The linearity of $\cP$ corresponds to a specific choice of the
set of preparing operations $\{\Phi\}$ for a given $\omega$, but
the specification of particular conditions remain to be an open problem.

Another open question is the characterization of all transformations
(not only with linear preparation map $\cP$) that can be understood 
within this model. This is indeed a very 
interesting, but also very difficult problem 
and we are not going to discuss this
here. Our aim is to describe the idea how to test
the compatibility of the original preparator and the action 
of an unknown device. The compatibility means that 
either the initial correlations vanish ($\Gamma=0$), 
or the unitary transformation generating the process dynamics
is from the set of transformations 
$U\in\{U_A\otimes U_B,U_A\otimes U_B U_{\rm SWAP}\}$. 
The pure state preparators are specific
examples of preparators with vanishing initial correlations ($\Gamma=0$),
and using the described procedure the pure state 
preparators are always compatible with 
an arbitrary quantum process. 

For instance, consider the preparator of a single pure state
described in the example of the perfect NOT realization.
That is, consider initially the spin is maximally entangled
with another spin and by measuring along the $z$ direction
we are preparing the states $|\uparrow\r$. Instead of
using different measurement apparatuses to generate
arbitrary pure states, let us perform single qubit rotations
of the state $|\uparrow\r$ to create arbitrary pure state
$|\psi\r$. As before, the device just swaps the two spins, i.e.
arbitrary $|\psi\r$ is replaced by $|\downarrow\r$. Consequently,
the transformation we obtain in the single point 
contraction of the Bloch sphere, i.e. linear and completely positive map
$\cE[\varrho]=|\downarrow\r\l\downarrow |$. That is, the same
black box can be described by different quantum processes depending on
the properties of the preparation procedures.

To test the dynamical compatibility of an unknown preparation device 
we suggest to exploit calibrated devices performing some well known 
quantum operations. In fact, this is nothing new, because 
the same procedure is used in most of the experiments. 
Important point is that using such method the properties of the 
resulting preparation map $\cP$ (and consequently $\cE$) 
strongly depends on the properties of
the original preparator. If one observes some ``unphysicality'' of $\cE$
in such experimental setting then before applying ``statistical 
corrections'' one should verify the compatibility of the preparator.
Performing the process tomography experiment 
the set of preparing operations should be chosen 
in a way that the generated set of states is sufficient for the process 
tomography, i.e. this set is finite. The goal of the verification is to
find whether the correlation matrix $\Gamma$ vanishes, or not.
There are, in principle, two strategies that can be combined. 
We can use different preparing operations $\Phi_1,\Phi_2$ generating
the same state $\varrho_{\Psi_1}=\varrho_{\Phi_2}$ and see whether the channel
action generates the same state. Alternatively, we can use additional
preparing operations to verify the linearity. In order to see, whether
the unphysicality is due to imperfections in the preparation process,
a characterization of all preparation maps $\cP$ in the described
settings is needed. This characterization is an open problem the solution 
of which is necessary if we want to be able to propose some universal 
verification and estimation strategies. 

\section{Evolution map for arbitrary time interval}

Let us assume that the time evolution 
of the system is described by a set of completely positive maps
$\cE_t$ induced by some underlying unitary dynamics $U_t$ 
of the system and its environment. This corresponds to a
situation in which initially (time $t=0$) the system and the environment are
factorized, i.e. $\omega_0=\varrho\otimes\xi$. Only under such 
assumption the maps $\cE_t$ can be completely positive for all $t$
and we have $\cE_t=\T U_t\varrho\otimes\xi U_t^\dagger$.
In this section we turn back to the original question posed 
at the beginning: what are the properties of the time 
evolution maps $\cE_{t_1,t_2}$ ($t_2>t_1$) describing 
the dynamics during arbitrary time interval $[t_1,t_2]$?

A direct calculation gives us that 
\be
\label{interevo}
\cE_{t_1,t_2}[\varrho]=\cE_{t_2}\circ\cE^{-1}_{t_1}[\varrho]\, .
\ee
These maps are linear, tracepreserving and hermiticity preserving, 
and thus defined on any operator (quantum state), 
but except $\varrho\in\cS=\cE_{t_1}[\cS(\cH)]\subset\cS(\cH)$ 
they can transform quantum states into negative operators.
Following the notation of Refs.\cite{jordan,shaji,shaji_dis,carteret} we define
the positivity domain $\cD_{\rm pos}(\cE)$ of a linear map $\cE$ as 
a subspace of states on which $\cE$ is positive, i.e.
$\cD_{\rm pos}(\cE)=\{\varrho\in\cS(\cH):\cE[\varrho]\ge 0\}$.
In fact, physically, only the action on such subset is important, 
because this is what we can really test in our experiments. 
Let us note that $\cE_{t_1}^{-1}$ corresponds to the mathematical inverse
operation (not physical) and does not necessarily always exist.
This means that the question about the form of $\cE_{t_1,t_2}$
does not make sense if $t_1$ is a point in which
the maps $\cE_{t_1}$ is not invertible.

The described model illustrates another physical situation,
in which the evolution map extended to whole state space 
is not completely positive. But in this case
it is still linear, what makes its description much simpler. 
Moreover, since the evolution maps $\cE_{t_1,t_2}$ preserves 
the trace and hermiticity, they can be expressed as a 
difference of two completely positive maps \cite{jordan,yu}
\be
\label{newform}
\cE_{t_1,t_2}[\varrho]=\sum_{j=1}^q A_j\varrho A_j^\dagger -\sum_{j=q+1}^{d^2} 
A_j\varrho A_j^\dagger\, ,
\ee
where the operators $A_j$ can be chosen so that they form 
an orthogonal operator basis ($\T A_j^\dagger A_k=0$ for $j\ne k$) 
and $d$ is the Hilbert space dimension of the system. 
The tracepreservation is reflected by
the identity $\sum_{j=1}^q A^\dagger_j A_j
-\sum_{j=q+1}^{d^2}A_j^\dagger A_j=I$.

The physics behind such form of noncomplete positivity is simple 
and just reflects the fact that at time $t_1$ the system is 
correlated to relevant degrees of freedom that affects the 
forthcoming evolution. In each time instance $t$ the global 
state of the system plus environment is described by some 
$\omega_t=U_t\omega_0 U_{-t}$. The preparation map 
in time $t$ is determined by the choice of $\omega_0=\varrho\otimes\xi$. 
In particular, $\cP_t[\varrho_t]=\omega_t=U_t\omega_0 U_{-t}
=U_t\cP_0[\varrho_0]U_{-t}=U_t \cP_0[\cE^{-1}[\varrho_t]] U_{-t}$, i.e.
$\cP_t=\cU_t\circ \cP_0\circ\cE_{t}^{-1}$. That is, the evolution for 
time interval $(t,t+\delta t)$ can be written as follows
\be
\cE_{\delta t}[\varrho_{t}]=
\T_B[U_{\delta t}\cP_{t}[\varrho_t]U^\dagger_{\delta t}]=\cE_{t+\delta t}\circ\cE_t^{-1}[\varrho_t]\, .
\ee
Let us note that we can generalize the whole setting by allowing
arbitrary preparation map $\cP_0$, but we want to preserve
the physical picture with the factorized preparation and 
therefore we will restrict ourselves to linear and factorized 
initial preparations only.

Before we get further let us note that the problem 
of quantum process tomography for linear noncompletely positive maps
was discussed in dissertation of Anil Shaji \cite{shaji_dis}.
To our knowledge this was the first attempt to understand 
the observed data in more general settings than just
in the framework of completely positive maps. 
In particular, he analyzes the problem from the mathematical 
point of view. Since the maps of this form are linear, the reconstruction
schemes based on complete data are the same independently 
whether the complete positivity constraint is applied, or not.
The inverse estimation schemes use just the linearity of the quantum 
evolution and the observed data are collected so that the 
result is represented by some linear transformation uniquely. The 
only and crucial question is what type of linear transformation 
it is. According to usual model of open system dynamics we expect
to obtain completely positive transformations in our experiments, 
but in reality this is not always the case \cite{howard,wunderlich}. 
Of course, the origin of this phenomena is questionable, 
but correlations of preparation map within the discussed model 
provide one possible option. The estimation schemes and 
algorithms must be modified accordingly in cases, 
when indirect statistical methods (such as maximum likelihood, 
or Bayesian approach) are employed, or our information is incomplete. 
Those who are interested in the details of complete quantum process 
tomography for noncompletely positive, but linear maps 
we refer to \cite{shaji_dis}. In this paper we are proposing 
the corresponding physical situation specifying the conditions
under which the linear noncompletely positive maps can be observed
experimentally.

Our interest is not only to perform the process tomography
experiment and reconstruction, but also to understand 
(at least partially) the physics behind. As we said 
the evolutions derived from the unitary 
dynamics $U_t$ for a fixed time interval are linear, tracepreserving and 
also hermiticity preserving. Therefore, they are of the 
form as written in Eq.(\ref{newform}). We are interested in the inverse
question, whether any such transformation $\cE$ can be understand as
subdynamics between two instants of time induced by some
unitary dynamics. Or alternatively, under which
circumstances the preparation mapping is linear on subset
of quantum states and whether these situations can be always 
understand as part of the dynamics described by $\cE_t$ 
derived from unitary dynamics $U_t$. 

If we assume that the one-parametric family of unitaries $U_t$ 
is arbitrary (i.e. the generating Hamiltonian is highly 
time dependent), then there are no constraints on the choice of 
the transformations $\cE_t$ for different times.
One can always define the generating unitary transformations
$U_{t}$ so that $\cT_B\circ\cU_t\circ\cP_0=\cE_t$.
Thus the question is, whether the following identity
can be fulfilled $\cE_{\delta t}=\cE_{t+\delta t}\circ\cE_t^{-1}$, where
on the left hand side we have arbitrary linear transformation
given by Eq.(\ref{newform}) and on the right hand side
we have two arbitrary completely positive maps $\cE_1,\cE_2$.
The inverse operation $\cE_t^{-1}$ is linear, tracepreserving and
hermiticity preserving as well, i.e. it serves as the potential 
source of noncomplete positivity. It is a well known fact that
substraction of two completely positive maps realizes arbitrary
noncompletely positive linear map (Eq.(\ref{newform})), 
but here the question is whether a similar property 
holds for the ``division'' of two completely
positive maps, i.e. for the transformation 
$\cE_{t+\delta t}\circ\cE_t^{-1}$. 

Consider now the following example. The transposition $\cE_{\rm trans}$ 
is probably the best known noncompletely positive linear map. Its action
is defined as follows $\cE_{\rm trans}[\varrho]=\varrho^T$ and for 
qubit it is closely related to perfect NOT operation. Let us assume 
that as the result of the process tomography we obtain the NOT operation,
$\cE_{\rm NOT}[\varrho]=\frac{1}{2}(\sigma_x\varrho\sigma_x
+\sigma_y\varrho\sigma_y+\sigma_z\varrho\sigma_z-\varrho)$. Could it 
happen within the discussed framework? Both these maps 
are positive, i.e. the positivity domain equals to the whole 
state space. In our settings the positivity domain always corresponds 
to an image of the whole state space under some completely 
positive map $\cE_t$. However, only for unitary transformations 
the image of the state space
equals to its original, i.e. if $\cE_t$ is a linear completely positive
map and $\cE_t[\cS(\cH)]=\cS(\cH)$, then $\cE_t$ is unitary. 
Therefore we have $\cE_t=\cU$. Because of the unitarity
this process is invertible. Consequently,
$\cE_{t+\delta t}\circ\cE_t^{-1}=\cE_{t+\delta t}\circ\cU^{-1}
=\cE_{\delta t}$ is necessarily a completely positive map. 
But this is in contradiction with the fact that our reconstruction 
gives us a noncompletely positive linear map $\cE_{\rm NOT}$ 
(or $\cE_{\rm trans}$). It means that the NOT operation 
$\cE_{\rm NOT}$, or transposition $\cE_{\rm trans}$
cannot be interpreted as an evolution map describing 
the time dynamics between two instants of time generated 
by a global unitary dynamics with initially factorized preparation map. 
In fact, the same conclusion holds for arbitrary positive (but 
not completely positive) linear map transforming pure states 
onto pure states, i.e. whenever the identity 
$\cE[\cS(\cH)]=\cS(\cH)$ holds.

However, there is still an option how to "partially'' perform the 
perfect NOT operation in the given framework 
of intermediate dynamics. The depolarizing single qubit 
channels form a one-parametric family
$\cE_{\{x\}}:\vec{r}\to x\vec{r}$, where
$\vec{r}$ is the Bloch vector corresponding
to a quantum state $\varrho=\frac{1}{2}(I+\vec{r}\cdot\vec{\sigma})$.
For $x\in[-1/3,1]$ the transformations $\cE_{\{x\}}$ are completely positive 
and $\cE_{\rm NOT}=\cE_{\{x=-1\}}$. For the inverse operations 
we have $\cE_{\{x\}}^{-1}=\cE_{\{1/x\}}$ and for the composition 
$\cE_{\{y\}}\circ\cE_{\{x\}}=\cE_{\{x.y\}}$ for arbitrary real $x,y$. Using
all these identities it simple to proof that 
$\cE_{\rm NOT}=\cE_{\{x\}}\circ \cE^{-1}_{\{-x\}}
=\cE_{\{x.(-1/x)\}}=\cE_{\{-1\}}$. This identity 
makes sense only for $-1/3\le x\le 1/3$, when
both transformations are completely positive. Formally, the above
calculation suggests that we are able to 
realize the perfect quantum NOT operation during the dynamics governed
by one-parametric set of completely positive maps. But, the 
above decomposition possessed a physical meaning only 
for states from the subset $\cE_{\{x\}}[\cS(\cH)]$, i.e. for states
with Bloch vectors smaller than $|x|$ ($|\vec{r}|\le |x|$). The maximal
set of states on which we are able to realize the perfect NOT 
operation (in the given model) is contained in the sphere with 
radius $x=1/3$. The conclusion is that the
perfect NOT operation can be find out as a result
of the process reconstruction. Moreover, it can be even understood as
an intermediate dynamics, but it does not mean that the perfect 
NOT operation is indeed accomplished, because
no intermediate dynamics can perform a perfect NOT operation 
for all set of states. Thus, process estimation of the quantum
operation between two instants of time could result in perfect
NOT operation. But the perfect NOT operation is performed only
on restricted set of states.

Although the presented framework of open system dynamics
enables us to explain quite naturally the experimental 
evidence of linear noncompletely positive maps,
the answer to the inverse question is open, i.e.
whether all hermiticity preserving, tracepreserving and linear 
transformations can be interpreted within the described model, if
we relax its physical validity for the whole set of states.
The characterisation of those noncompletely positive maps that 
can be realized within the discussed physical model 
is an open question that indeed requires deeper investigation.

\section{Conclusion}

In this paper we have analyzed the dynamics of open 
quantum systems beyond the complete positivity restriction 
and related consequences for the
process tomography. The correlations
can be detected if the direct estimation procedure gives
physically invalid result and simultaneously, all the experimental 
and statistical deviations can be excluded. The good news of our analysis
is that any failure of the process estimation (i.e. "unphysical'' result) 
cannot be interpreted as the failure of quantum theory unless one
can exclude the presence of initial correlations and 
dynamically incompatible preparators. However, the
particular realization using the SWAP operation is 
quite artificial. As an example, we have described
two different experimental realizations of the perfect NOT gate, 
which is considered
to be unphysical. The bad news is that although the initial
correlations can be detected, the process tomography is very 
difficult and ambiguous, and the physical origin 
could be quite artificial. However, the usage of statistical 
tools is justified only if we can safely exclude all such
(artificial) possibilities, i.e. we have some restrictions and models
on the form of possible preparator maps.

We have discussed two very
natural physical situations that can result in observation
of initial correlations effect. First of them is motivated by current
experiments, in which experimenters typically use single state
preparator and other states are generated with the help
of further processing, for instance by applying different 
unitary operations. In the second case the ``unphysicality''
is related to the fact that the evolution map
describing the state dynamics during arbitrary time interval
is in general not completely positive, but still it is linear.
This situation is very closely related to experiments
in process tomography, in which the state estimation of inputs
is as necessary as the state estimation of the outputs, i.e.
we indeed perform measurements in two time instants. And it could
happen that already at the first time instant the system's
and the channel's degrees of freedom are mutually (although weakly) 
correlated. Fortunately, from the practical point of view,
in this case the process tomography schemes are not affected, 
only the data processing should be different if one uses 
some advanced statistical tools. We think that this 
framework provides a physically reasonable description for
the existence of noncompletely positive linear maps. 
We have shown that in a strict sense 
not all linear noncompletely positive maps can be 
indeed realized within such model. The question of the
characterization of ``accessible'' maps within this model 
is interesting and very important, but unfortunately we do not 
know the answer yet.

We have argued that the problem of initial correlations 
is not a problem of quantum dynamics, but
rather of quantum kinematics. In other words, the process tomography 
always describes a relation between the preparators and the channel.
For different sets of preparators physically the same channel could
be described by different dynamical maps. Only in very specific 
cases of preparators the channel is represented
as a completely positive tracepreserving linear map.
Fortunately, this is the case that usually holds in labs, 
or better to say, we are aiming to hold in our labs. 
We have described the method how to test the preparators 
using the calibrated quantum channels. Good preparator devices
are crucial for the successful development of quantum 
information processing. The presented analysis
is very far from being complete and a deeper investigation
on the system-environment correlations effects on quantum dynamics and
experiments is needed.


\section{Acknowledgements} 
I would like to thank Peter \v Stelmachovi\v c and Jason
Twamley for discussion. This work was supported in part by the European
Union  projects QAP 2004-IST-FETPI-15848 and 
CONQUEST MRTN-CT-2003-505089, by the Slovak
Academy of Sciences via the project CE-PI, by the project
APVT-99-012304, VEGA 2/6070/26 and GA\v CR GA201/01/0413.



\begin{thebibliography}{10}
\bibitem{note1}
In Refs.\cite{jordan,shaji_dis} this map is called an assignment map, 
in Ref.\cite{carteret} it is called an extension map, but here we will
use the name preparation map expressing our motivation.
\bibitem{access_note}{Let us note a difference
between the notion of accessible map in Ref.\cite{carteret}
and  its usage in this paper. In \cite{carteret} the accessible maps
are those for which the preparation map is linear.}
\bibitem{nonphysicality}
In most of the current experiments realizing 
quantum gates, the presented operations 
(obtained from the direct estimation method) 
are not always completely positive. The problems 
related to occurence of noncompletely positive maps 
in experiments were analyzed in Refs.\cite{boulant,weinstein,howard}.
\end{thebibliography}
\end{document}